# Functional light diffusers based on hybrid CsPbBr$_3$/SiO$_2$ aero-framework structures for laser light illumination and conversion

Lena M. Saure[1], Jonas Lumma[1,2], Niklas Kohlmann[3], Torge Hartig[4], Ercules E.S. Teotonio[2,5], Shwetha Shetty[6], Narayanan Ravishankar[6], Lorenz Kienle[3,7], Franz Faupel[4], Stefan Schröder[4], Rainer Adelung[1,7], Huayna Terraschke[2,7*], Fabian Schütt[1,7*]

[1] Functional Nanomaterials, Department for Materials Science, Kiel University, Kaiser Str. 2, 24143 Kiel, Germany

[2] Institute of Inorganic Chemistry, Christian-Albrechts-Universität zu Kiel, Max-Eyth-Str. 2, D-24118 Kiel, Germany. E-Mail: hterraschke@ac.uni-kiel.de

[3] Synthesis and Real Structure, Institute for Materials Science, Kiel University, Kaiser Str. 2, 24143 Kiel, Germany

[4] Chair for Multicomponent Materials, Department for Materials Science, Kiel University, Kaiser Str. 2, 24143 Kiel, Germany

[5] Department of Chemistry, Federal University of Paraíba, 58051-970 João Pessoa, Paraíba, Brazil

[6] Materials Research Centre, Indian Institute of Science, Bangalore, India

[7] Kiel Nano, Surface and Interface Science KiNSIS, Kiel University, Christian-Albrechts-Platz 4, 24118 Kiel, Germany. E-Mail: fas@tf.uni-kiel.de

# Abstract

The new generation of laser-based solid-state lighting (SSL) white light sources requires new material systems capable of withstanding, diffusing and converting high intensity laser light. State-of-the-art systems use a blue light emitting diode (LED) or laser diode (LD) in combination with color conversion materials, such as yellow emitting Ce-doped phosphors or red and green emitting quantum dots (QD), to produce white light. However, for laser-based high-brightness illumination in particular, thermal management is a major challenge, and in addition, a light diffuser is required to diffuse the highly focused laser beam. Here, we present a hybrid material system that simultaneously enables efficient, uniform light distribution and color conversion of a blue LD, while ensuring good thermal management even at high laser powers of up to 5W. A highly open porous (> 99%) framework structure of hollow $SiO_2$ microtubes is utilized as an efficient light diffuser that can drastically reduce speckle contrast. By further functionalizing the microtubes with halide perovskite QDs ($SiO_2$@$CsPbBr_3$ as model system) color conversion from UV to visible light is achieved. Under laser illumination, the open porous structure prevents heat accumulation and thermal quenching of the QDs. By depositing an ultrathin (~ 5.5 nm) film of poly(ethylene glycol dimethyl acrylate) (pEGDMA) via initiated chemical vapor deposition (iCVD), the luminescent stability of the QDs against moisture is enhanced. The demonstrated hybrid material system paves the way for the design of advanced and functional laser light diffusers and converters that can meet the challenges associated with laser-based SSL applications.

# Introduction

In solid-state lighting (SSL) light emitting diodes (LEDs) are the most prominent candidates known for the generation of white light to replace traditional lighting technologies. However, with the demand for efficient high-brightness and high-power lighting applications, LEDs are limited in performance by the so-called 'efficiency droop', when reaching high input power densities.[1,2] Laser diodes (LDs) are an alternative SSL technology due to their high electro-optical conversion efficiency.[1,2] State-of-the-art systems for white light generation are blue LEDs or LDs in combination with a yellow emitting phosphor, e.g. Cerium doped yttrium aluminum garnet (YAG:Ce) phosphors to create white light by color mixing of blue and yellow.[1,3] While for LEDs good color rendering index (CRI) values with long-term stability can be achieved, laser-driven lighting often lacks stability due to heat accumulation by high power luminous flux, leading to thermal quenching of the phosphor.[1,3] Thus, especially for laser-based systems, thermal management is one of the major challenges for the next generation of SSL technology.[1,3]

Besides yellow emitting phosphors and related lanthanide-based materials that are used for laser-based lighting and projection, as well as in white light emitting diodes (WLED), quantum dots (QDs) gained high attention for color conversion in lighting devices such as quantum dot light emitting diodes (QLED) for low-cost light sources and display applications as well as projectors[4–9]. State-of-the-art systems utilize red and green emitting QDs in combination with a blue LED. One of the main challenges, especially of halide perovskite QDs, which exhibit extraordinary optoelectronic properties, remains their environmental stability against moisture, oxygen, UV light exposure and high temperatures.[4,5,10,11] Manifold strategies to protect the QDs have been developed including e.g. the fabrication of core-shell quantum dots with different ligand system, silica and polymer shells.[10] Also, encapsulation of QD solutions[12] has been reported for the application in LEDs, as well as embedding of QDs into silica monoliths[13] and aerogels[4,6,14]. Several approaches are based on the incorporation of QDs into hydrophobic silica aerogels, while maintaining the exeptional photoluminescence and improving the moisture stability[6]. QD-aerogel composite powders are then embedded into a polymer film as backlight display or mixed into a silicone resin to build WLEDs.[4,6] Homogeneous doping of silica aerogels with QDs was reported by Lazovski et al., thereby immobilizing the QDs and simultaneously maintaining high accessibility for e.g. catalysis applications.[14]

However, while all these studies demonstrate that aerogels are a very promising host material for QDs and can be used to create white light based on LEDs, two main challenges remain to be solved to enable laser-driven lighting applications: (1) Aerogels have very low light scattering properties[14–16] and are not suitable to diffuse the highly focused laser light uniformly in the room. Thus, an additional light diffuser is required. (2) Aerogels are known for good thermal insulation[17], hence, gas exchange in and out of the structure is limited and could lead to heat accumulation and thermal quenching of the QDs.

Recently, nanomaterial-based three-dimensional (3D) porous foam structures have gained attention as light diffusers for laser light, showing scattering efficiencies of > 98 % while simultaneously reducing the speckle contrast.[18,19] Using a red, green and blue (RGB-) laser system, laser-based white light illumination without phosphorous materials was demonstrated.[18,19] However, using three different LD drastically increases the costs of the final device and, up to now, green laser diodes still suffer from poor efficiencies, limiting the overall performance of the device. [5,20]

To address the challenges of laser-based lighting applications, we have designed a highly porous (> 99%) hybrid material system based on a $SiO_2$ framework structure, named aeroglass, which efficiently scatters light in all spatial angles and acts as a light diffuser that can reduce the speckle contrast of laser light to a value of ~5.85%. The combination of aeroglass with halide perovskite $CsPBr_3$ QDs as a model system enables color conversion for white light emitting devices. The open porous structure of aeroglass combined with its low heat capacity prevents heat accumulation and thermal degradation of the QDs. To further improve the moisture stability of the hybrid material system, we show that the addition of an ultra-thin film (~5.5 nm) of polyethylene glycol dimethylacrylate (pEGDMA) via an initial CVD (iCVD) process on the entire 3D hybrid framework structures results in a globally hydrophobic structure that significantly increases the water stability of the hybrid system compared to the non-coated structure.

# Results

**Figure 1** provides a direct comparison between conventional silica aerogels and the here demonstrated $SiO_2$ framework structures. While the aerogel is characterized by a translucent appearance (**Figure 1**a), the aeroglass appears whitish (**Figure 1**b), even though its density (~3 mg cm$^{-3}$) is lower compared to classical aerogels (150 mg cm$^{-3}$). The difference in optical appearance is a direct result of the different micro- and nanostructure. SEM images (**Figure 1**c-f) clearly show the structural difference between aerogels and aeroglass. Silica aerogels are composed of nm-scale $SiO_2$ particles framework (**Figure 1**c, e), whereas aeroglass is composed of widely interconnected hollow $SiO_2$ microtubes with nanoscale wall thickness (**Figure 1**d, f).

Due to their structure, silica aerogels are optically homogeneous materials as the wavelength of light in the visible spectrum ($\lambda_{vis}$) is much larger than the individual $SiO_2$ particles forming the aerogel.[18] In contrast, aeroglass has a hierarchical microstructure with different sized features compared to $\lambda_{vis}$ (see **Figure S1**). It consists of a network of hollow microtubes (diameter of 1-3 μm) with nanoscopic wall thickness (~ 17 nm) and distance of up to 100-300 μm between individual microtubes, acting as randomly distributed Rayleigh scattering centers that result in diffusive light scattering. More details are described by Schütt et al. for a similar structure composed of hexagonal boron nitride[19].

The difference in optical properties can be seen when the materials are placed on top of a laser pointer (**Figure 1**g, h). The silica aerogel transmits most of the light with only a few scattering events due to defects in the aerogel structure. In contrast, aeroglass scatters incident light in all spatial angles (**Figure 1**h).

The fabrication process is schematically depicted in **Figure 1**i. In brief, a network of interconnected ZnO microparticles (density 0.3 g cm$^{-3}$) with tetrapodal shape is wet-chemically coated with a thin layer of $SiO_2$ based on the Stoeber synthesis.[21,22] Etching of the sacrificial ZnO template followed by supercritical drying results in a freestanding network of interconnected hollow $SiO_2$ microtubes (details in materials and methods) with a density of ~3 mg cm$^{-3}$. The density of $SiO_2$ microtubes is calculated to be ~6.8*10$^8$ cm$^{-3}$ (see SI Note 1). Tailoring the density of the sacrificial ZnO template, e.g. to 0.6 g cm$^{-3}$ or 0.9 g cm$^{-3}$, directly influences the density of the hollow microtubes, i.e. the scattering centers, within the aeroglass structure (~13.6*10$^8$ cm$^{-3}$ or ~20.4*10$^8$ cm$^{-3}$, respectively). Additional SEM images are shown in **Figure S2**. In addition, the synthesis method allows the functionalization of the microtube arms with nanoparticles, which extends the properties of the light scattering framework structure.

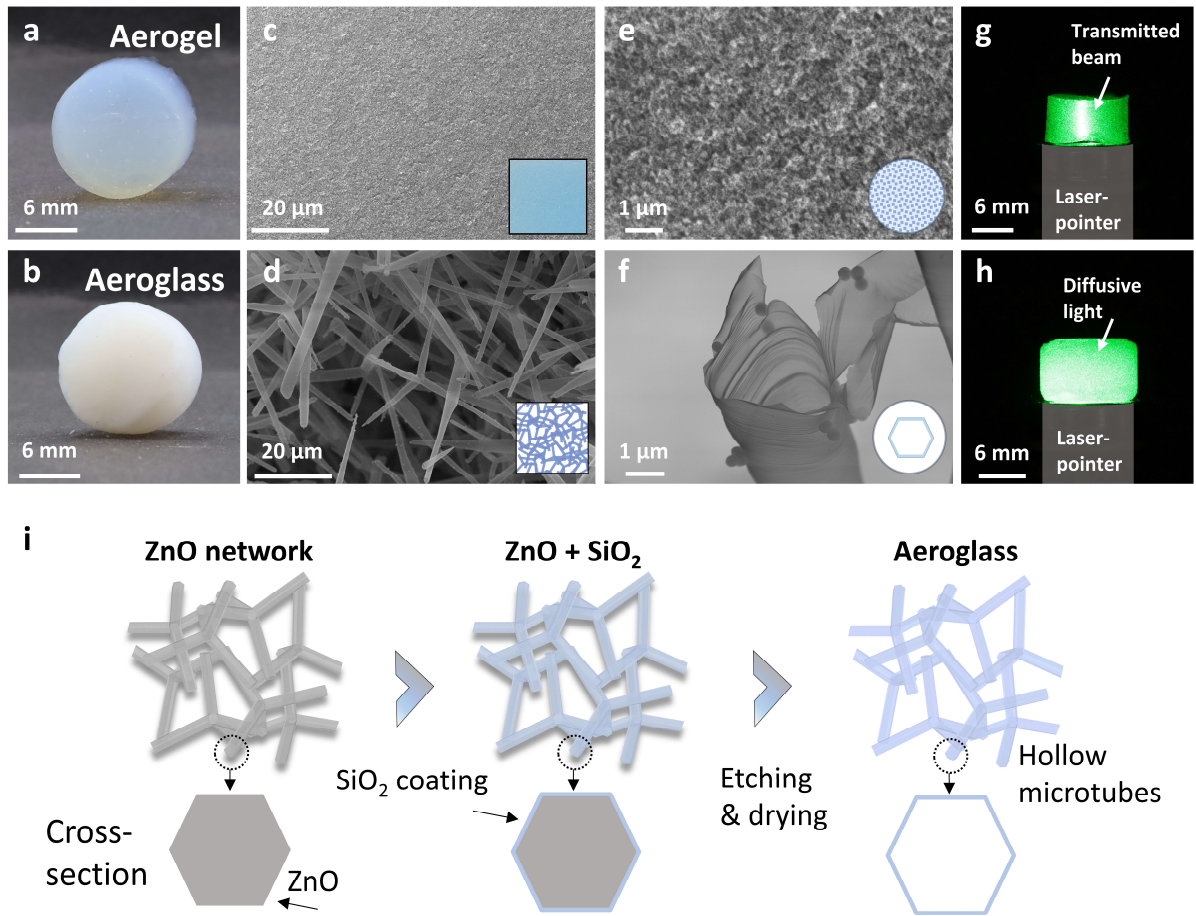

**Figure 1** (a) Photograph of silica aerogel and (b) aeroglass under white light room illumination. (c)-(f) SEM images of (c), (e) silica aerogel and (d), (f) aeroglass with two magnifications revealing the difference in the microstructure. (g), (h) Photograph of silica aerogel and aeroglass on top of a green laser pointer. (i) Schematic of fabrication process of aeroglass.

The characteristic light scattering properties in the x-y-plane and y-z-plane, respectively, of a classical silica aerogel and aeroglass with different densities of scattering centers (~$6.8*10^8$ cm$^{-3}$, ~$13.6*10^8$ cm$^{-3}$ and ~$20.4*10^8$ cm$^{-3}$, in the following referred to as low, medium and high density of microtubes per unit volume) measured with a photogoniometer are shown in **Figure 2**a, b. For the aerogel (density of ~150 mg cm$^{-3}$), only a transmitting beam can be detected, while different microtube densities within the aeroglass show homogenous light scattering in all spatial angles. Aeroglass with low density of microtubes shows more forward scattering compared to aeroglass with medium density of microtubes, which shows a uniform light distribution in all spatial directions. For a high density of microtubes backward scattering is more prominent (**Figure 2**b). The spatial light scattering in the x-z-plane can be found in the supporting information, **Figure S3**.

**Figure 2**c and d demonstrate the room illumination using a 5W blue laser. Using solely a blue laser, a bright spot appears on the wall, whereas mounting an aeroglass with low density of scattering centers in front of the laser, light is homogenously illuminating the room. In addition,

**Figure S**4 demonstrates the speckle reduction capability of aeroglass. A speckle contrast of 5.85 % ± 1.03 % was calculated (details SI Note 2), which is similar compared to speckle reduction by other nanomaterial-based foam structures.[18,19]

The ability to withstand high laser power densities without accumulation of heat is demonstrated by infrared (IR) thermography (**Figure 2**e). The aeroglass sample heats up to only ~39 °C within 1 min of illumination with a 5W-blue laser, as the open porous structure enables efficient gas flow in and out of the structure. Photographs before and after illumination show no effect on the $SiO_2$ framework structure (**Figure S7**).

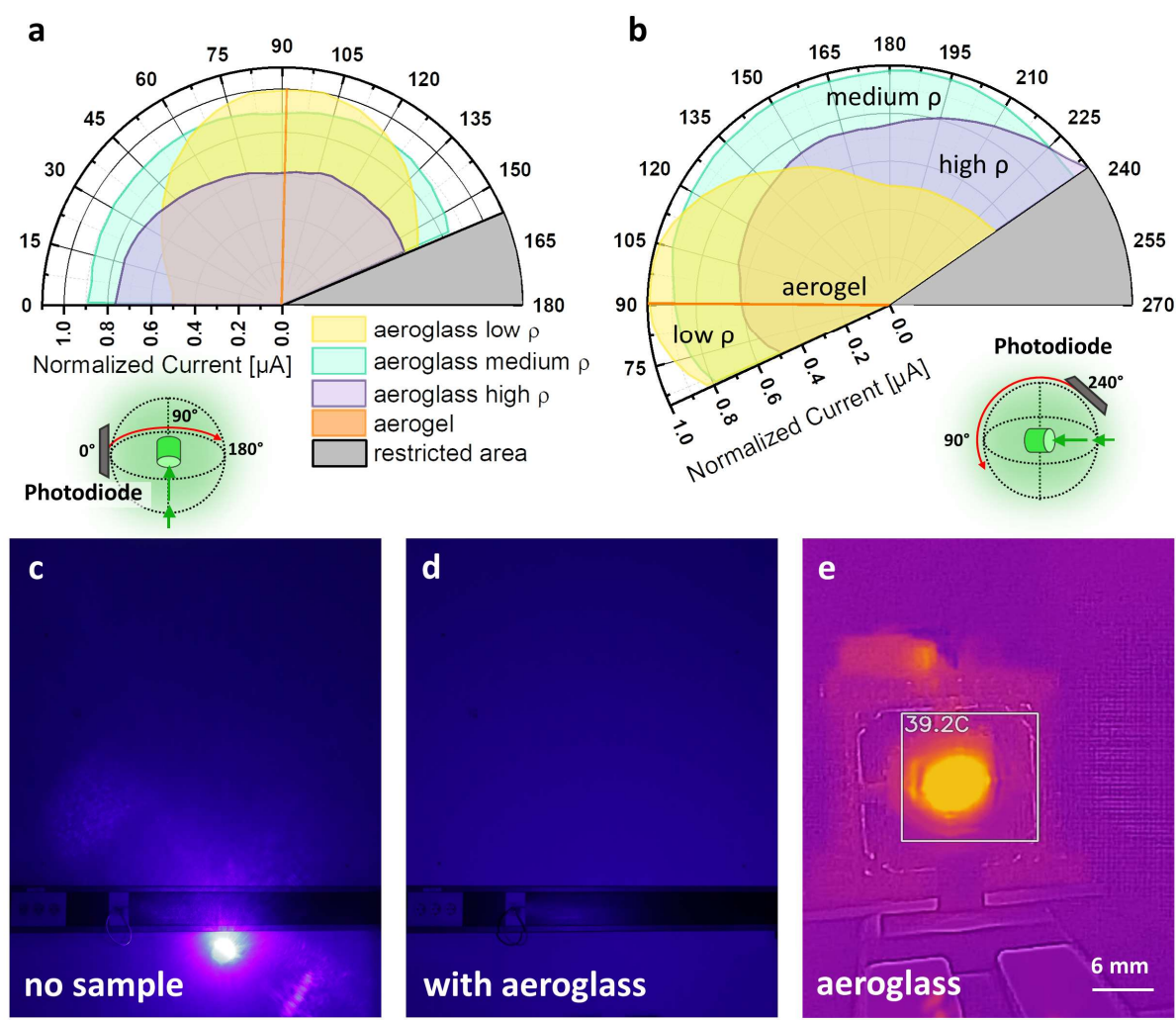

**Figure 2** Light scattering of silica aerogel and aeroglass with low, medium and high density of scattering centers measured with a photogoniometer. (a) Scattering intensity of x-y-plane and (b) in z-plane. (c) Room illumination using a 5W-blue laser without any light diffuser and (d) with an aeroglass sample with low density placed the laser beam. (e) thermograph of an aeroglass illuminated with a 5W-blue laser for 1 min.

**Functionalization with halide perovskite QDs**

The combination of the light scattering properties with the effective speckle reduction can be further extended to color conversion properties by functionalizing the $SiO_2$ framework structure with color converting materials (see **Figure 3**a). Here, inorganic halide perovskite core shell QDs ($SiO_2$@$CsPbBr_3$) based on the synthesis route of Zhang et al.[23] are synthesized as a model system. Prior fabricated aeroglass samples are immersed in the QD solution during the growth of the $SiO_2$ shell (see **Figure 3**a). Note, that for the functionalization aeroglass samples based on a ZnO template density of 0.3 g cm$^{-3}$ have been used. The QD functionalized aeroglass hybrid framework structures are henceforth referred to as aeroglass-QD. SEM and TEM images (**Figure 3**b-d) obtained after washing and supercritical drying of aeroglass-QD samples reveal a homogeneous distribution of QDs on the surface of the hollow $SiO_2$ microtubes. EDX mapping (**Figure 3**e) confirms the formation of $SiO_2$@$CsPbBr_3$ QDs by the presence of the corresponding chemical elements. In addition to the formation of single particles, larger particles (see **Figure 3**f) formed on the surface of the microtube arms, consisting of individual $SiO_2$@$CsPbBr_3$ particles with a size of ~5 nm (**Figure 3**g). By taking the rotational average followed by background subtraction via a polynomial model of the selected area diffraction (SAED) pattern containing numerous nanoparticles, the overall pattern can be assigned to cubic $CsPbBr_3$ (see **Figure 3**h), which is in accordance with Zhang et al.[23], whereas the underlying $SiO_2$ has an amorphous structure (see **Figure S5**). **Figure 3**i and j show photographs of a functionalized aeroglass sample before washing and drying under white light and UV illumination, respectively. The yellowish but still transparent character results from the attachment of QDs to the $SiO_2$ framework structure. Illumination with UV light (364 nm) leads to fluorescence of the QDs and emission of green light.

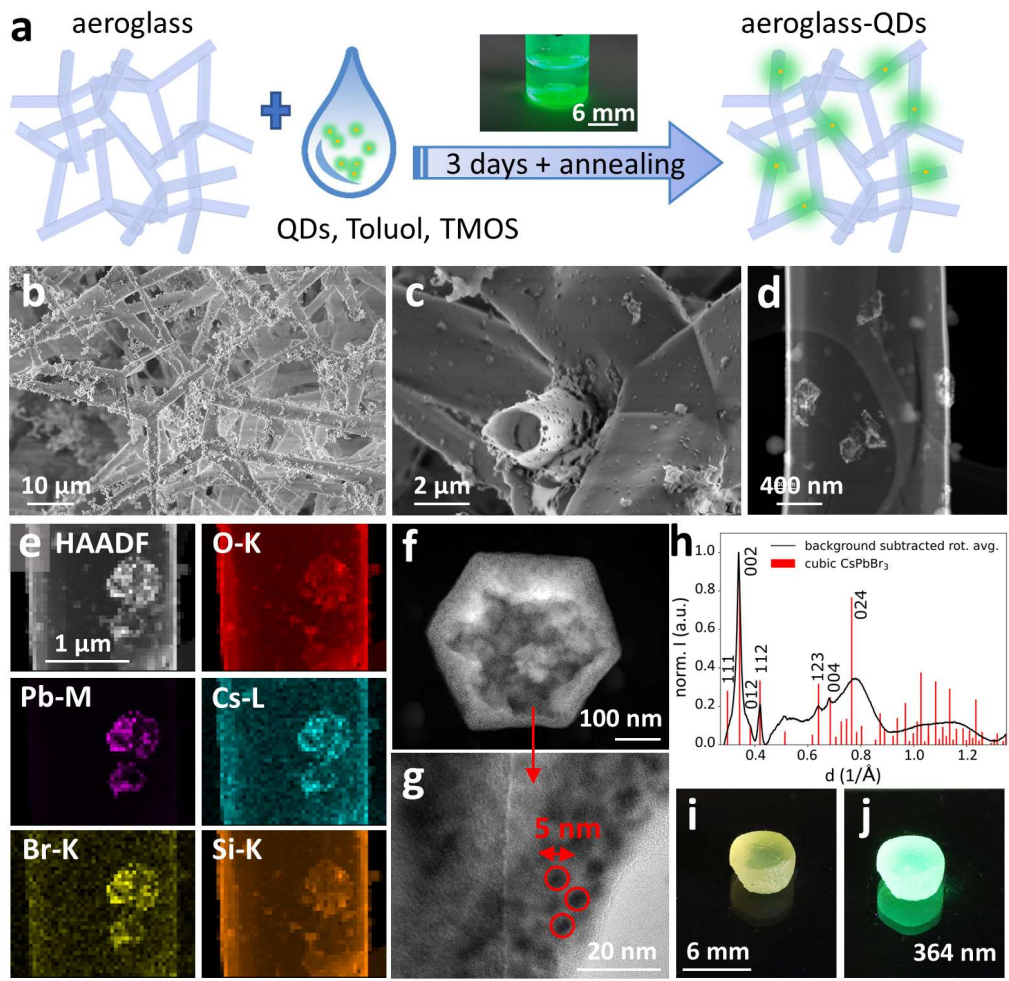

**Figure 3** (a) Schematic of functionalization process of aeroglass samples with QDs. (b), (c) SEM images and (d) TEM images of an aeroglass-QD hybrid sample. (e) EDX mapping of an aeroglass microtube functionalized with QDs. (f) TEM image of large particle consisting of (g) individual QDs. (h) Rotational average of SAED pattern of aeroglass-QD sample revealing a cubic phase of the $CsPbBr_3$ QDs. (i), (g) Photograph of aeroglass-QD sample prior to supercritical drying under white light and UV illumination, respectively.

**Polymer thin film coating via iCVD for improved moisture stability**

Due to the highly porous structure of aeroglass, water is immediately absorbed into the structure by capillary forces upon contact, causing the structure to collapse upon air drying. Furthermore, exposure to water can degrade the luminescence of QDs. Thus, to enable high moisture stability of the aeroglass-QD hybrid structures, we introduce an iCVD polymer coating step on the aeroglass-QD hybrid structure to achieve global hydrophobic properties, as shown in Figure 4, without losing the highly open porous character of the structure. Therefore, the as-synthesized aeroglass-QD hybrid structures were coated with an ultrathin layer (~5.5 nm) of poly ethylene glycol dimethylacrylate (pEGDMA).

TEM investigations confirm an increase of the relative microtube wall thickness (measured by EELS) in terms of effective inelastic electron mean free path (eMFP) from 0.233 eMFP to

0.306 eMFP (more details in SI Note 3). Further, an increase in carbon content (nominally from ~10 at% to ~24 at% by EDX, cf. **Figure 4**a, b) is indicating a homogeneous coating of pEGDMA on the microtubes. This is also shown in EDX mapping of an aeroglass-QD-iCVD microtube arm (**Figure 4**c).

Luminescence measurements of aeroglass-QD and aeroglass-QD-iCVD (**Figure 4**d, e) show the excitation (blue curves) assigned to the maxima of the respective emission spectra, recorded for an excitation wavelength of 275 nm (red curves). Both sample types show a similar maximum emission wavelength of 505 nm and 507 nm, respectively, while for pure aeroglass no luminescence is detected. Interestingly, the aeroglass-QD-iCVD sample shows an additional shoulder in the emission spectrum between 300 nm and 380 nm. This might be attributed to the pEGDMA coating. Very similar emission bands have been recently reported by X. Z. Kong et al.[24] for several polyethylene glycol derivatives and attributed to cluster formation of its chains and the presence of lone pairs of electrons in the heteroatoms.

**Figure 4**f-h show photographs of a pure aeroglass, an aeroglass-QD and an aeroglass-QD-iCVD sample illuminated with a blue laser pointer. Functionalization with QDs results in a change of color emitted from the aeroglass-QD sample compared to the pure aeroglass, as the $SiO_2$@$CsPbBr_3$ QDs partially convert the blue into green light. The effect of the iCVD coating on the water contact angle of aeroglass-QD samples is demonstrated in **Figure 4**i. For samples without additional coating, water droplets get soaked into the structure due to capillary forces (see SI Video 1 and **Figure 5**). This was reported before for similar 3D framework structures.[25,26] However, the additional pEGDMA coating results in a global hydrophobicity and the water droplet remains on top of the sample.[25,26]

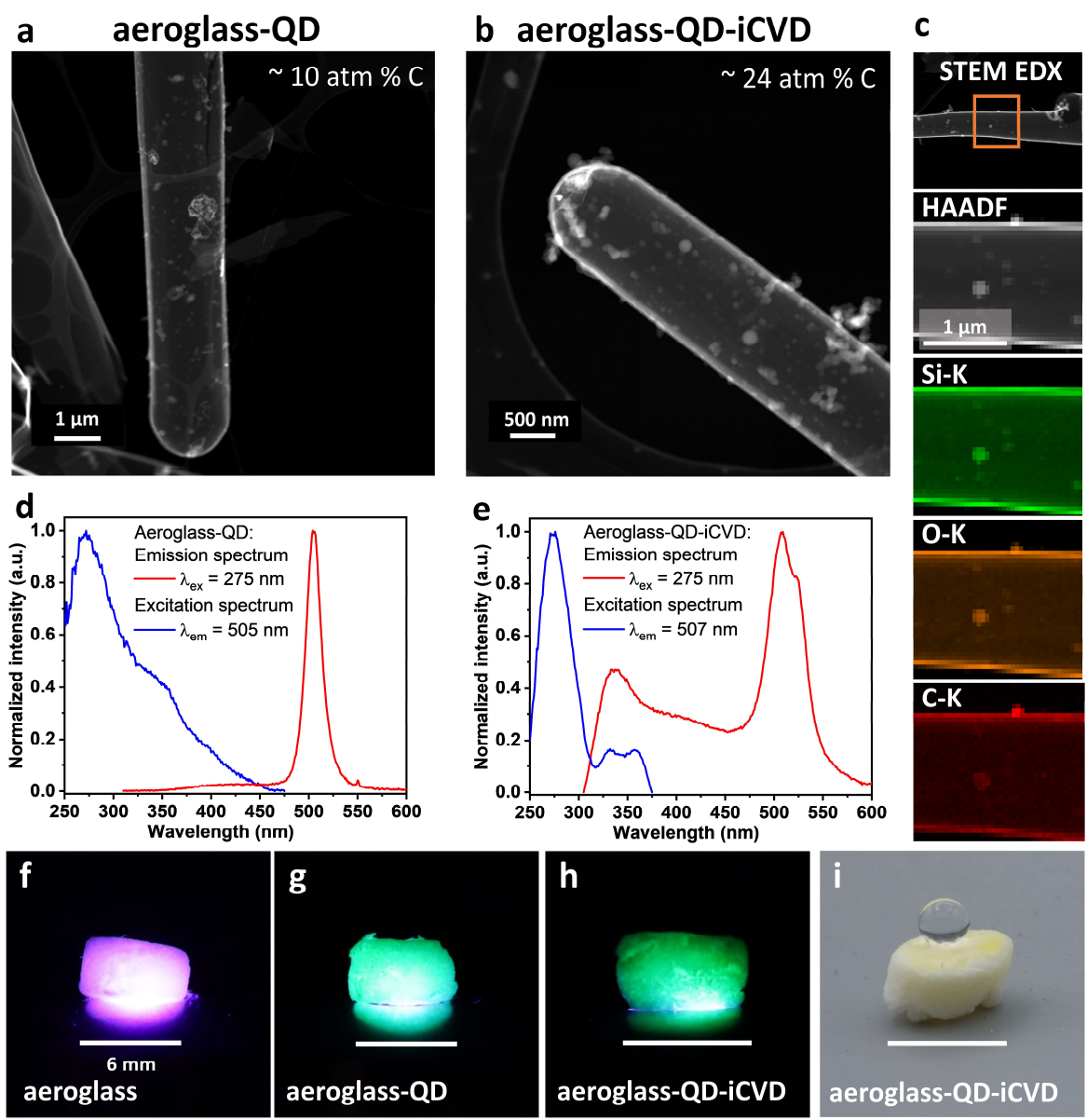

**Figure 4** STEM images of (a) aeroglass-QD and (b) aeroglass-QD-iCVD hybrid samples with measured carbon content. (c) EDX mapping of a microtube of an aeroglass-QD-iCVD hybrid sample. Luminescence of (d) aeroglass-QD and (e) aeroglass-QD-iCVD hybrid samples with excitation at 275 nm. (f)-(h) Photographs of aeroglass, aeroglass-QD and aeroglass-QD-iCVD, respectively, illuminated with a blue laser pointer. (i) Photograph of aeroglass-QD-iCVD sample with a water droplet on top of the sample, demonstrating the global hydrophobicity.

Long-term stability of the $SiO_2$@$CsPbBr_3$ QDs against moisture was investigated by immersing an aeroglass-QD sample as well as an aeroglass-QD-iCVD sample in water (see **Figure 5**a, b). Images were recorded under UV light (375 nm) at different time points. The free volume of pure aeroglass-QD samples is immediately filled with water due to high capillary forces, whereas the aeroglass-QD-iCVD sample is floating on top of the water due to its global hydrophobicity. It has to be noted that the pEGDMA coating does not change the morphology[26] of the aeroglass, as only a conformal ultra-thin film is deposited on the microtubes. The luminescence of the non-coated samples decreases within minutes of water contact. In contrast, the luminescence of the iCVD coated sample is significantly improved. It remains stable for more than 21 h and after 42 days the sample is still floating on top of the water showing a weak luminescence. Additional images in **Figure S6** show the gradual decrease of luminescence over time.

a Aeroglass-QD
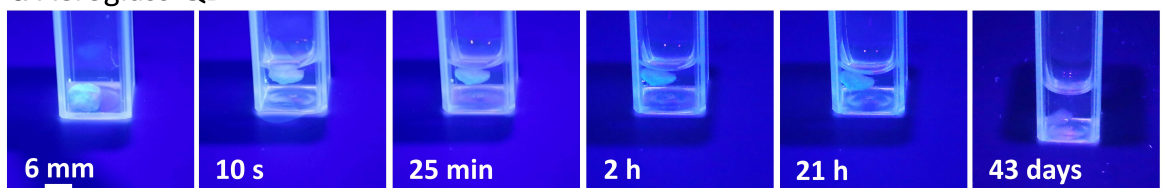
b Aeroglass-QD-iCVD
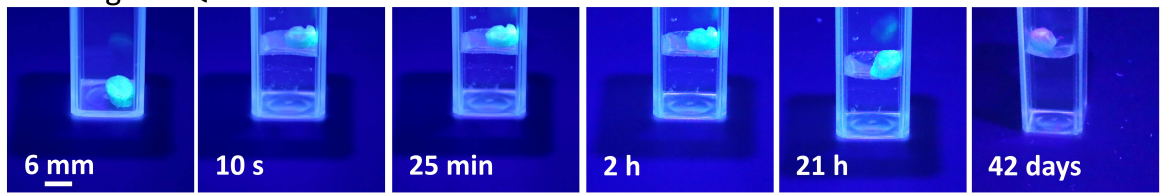

**Figure 5** Long-term water stability of (a) aeroglass-QD and (b) aeroglass-QD-iCVD samples, showing luminescence under UV illumination at different time points. (c) Thermograph of aeroglass after 1 min illumination with a 5W blue laser. (d) Room illumination with a 5W laser and (e) with an aeroglass sample mounted in front of the laser, resulting in uniform room illumination.

## Conclusion

In summary, we demonstrated a hybrid $SiO_2$ framework structure functionalized with halide perovskite QDs ($SiO_2$@$CsPbBr_3$) as a model system for laser-based white light generation using a simple wet-chemical approach. The hybrid material system presented here, efficiently scatters laser light with significantly reduced speckle contrast (~5.85%), can withstand high light power densities, and simultaneously acts as a color conversion material. The well-known moisture instability of QDs is addressed by applying an ultra-thin pEGDMA coating (~5.5 nm) via iCVD on the entire 3D framework structure, resulting in global hydrophobic properties and significantly increased water stability compared to the non-coated material. The open porous framework structure allows efficient gas exchange in and out of the structure, preventing heat accumulation and thermal quenching of the QDs in the material system. The fabrication approach can be easily extended to other nanomaterials, such as different types of QDs and catalytic particles, making the SiO$_2$ framework structure an ideal candidate for laser-based white light sources, e.g. using a blue LD in combination with red and green emitting QDs. An application in photocatalysis could also be envisioned, as the open porous structure allows efficient gas exchange and provides high gravimetric surface area, which could be combined with catalytic particles.

## Methods and Materials

**Silica aerogels** were fabricated based on a sol-gel synthesis route based on a recipe from *aerogel.org*. A stock solution is prepared using 1.852 g NH$_4$, 100 ml distilled water and 22.78 ml ammonium hydroxide solution. In a beaker, 5 ml of tetraethyl orthosilicate (TEOS) and 11 ml ethanol are mixed. In a second beaker, 7 ml water and 11 ml ethanol are mixed, followed by the addition of 0.371 ml stock solution. The second solution is poured into the first solution and stirred well, forming the sol, which is then poured into cylindrical molds of desired geometry. After the gel has formed, the molds are immersed in ethanol. After 24 h the samples are gently pressed out of the molds, and stored in ethanol for 4 more days to allow the gel to age. The ethanol is changed every day to remove all residuals of the precursors. In the last step, the aerogel samples are dried in a supercritical point dryer (Leica EM CPD300).

**Aeroglass** samples were fabricated using a wet-chemical approach based on sacrificial porous ceramic templates of tetrapodal ZnO (t-ZnO). Briefly, t-ZnO powder was produced in a flame transport synthesis, as reported elsewhere[27,28], pressed into templates of desired geometry and density, and sintered at 1150°C for 5 h to form interconnected t-ZnO networks[29,30]. Following, the templates were coated with a thin film of silicon dioxide (SiO$_2$) based on the Stoeber synthesis of SiO$_2$[21,22]. In more detail, ethanol, TEOS and ammonium hydroxide were mixed in a ratio of 10:0.1:3 and t-ZnO templates were immersed for 45 min, rinsed in ethanol and stored in water for 24 h. The sacrificial template was then removed using hydrochloric acid, followed by thorough washing with first distilled water and then ethanol, and drying with a supercritical point dryer. This results in highly porous networks of interconnected hollow SiO$_2$ microtubes, named aeroglass. For higher stability all samples intended for functionalization with quantum dots (QDs) were coated twice with SiO$_2$, before etching of ZnO was performed.

**Quantum Dots (QDs, CsPbBr$_3$)** were fabricated by adapting the protocol of Zhang et al.[23] First a precursor solution of CsBr (0.4 mmol) and PbBr$_2$ (0.2 mmol) in N,N-dimethylformamide was prepared. Oleic acid (0.5 ml) and Oleylamine (0.25 ml) were added to this solution. The quantum dots formed immediately by injecting 0.5 ml of the precursor solution into 2 ml toluene, which was stirred with 500 rpm. By centrifugation (15000 rpm) the quantum dots were separated from the reaction solution. The received product was re-dispersed in 2 ml toluene.

**Growing of a SiO$_2$ shell on the QDs and functionalization of aeroglass with QDs** was carried out according to the following protocol: An aeroglass sample was placed in 1 ml of the QD dispersion. Afterwards, 40 µl tetramethyl orthosilicate (TMOS) were added. The samples were stored at room temperature for 3 days, followed by annealing to 40 °C for 30 min. Toluene was replaced with ethanol, washed multiple times with absolute ethanol and dried with a critical point dryer to obtain dry aeroglass-QD samples.

**Light scattering** properties were measured with a self-built photogoniometer using a photodiode (FDS1010, Thorlabs) that is rotated in steps of 5° around the sample at a distance of ~ 15 cm. The samples were placed in the center of the device and were illuminated with an RGB laser module (RTI OEM 300 mW RGB Modul, LaserWorld). Each single laser has a maximum power of 100 mW and a focused spot size of ~1 mm. The samples had a cylindrical geometry with a diameter of 12 mm and a height of 10 mm and a conical taphole. To demonstrate laser-based room illumination a 5W laser module with a wavelength of 450 nm.

**Infrared thermography** was performed using a FLIRONE Pro infrared camera.

**SEM** measurements were performed using a Zeiss Supra 55VP.

**TEM** measurements were performed using a FEI Tecnai F30 G$^2$ STwin operated at 300 kV. The microscope is equipped with an EDAX Si/Li detector for elemental analysis via EDX and a Gatan Tridiem 863P post column image filter (GIF) allowing for electron energy loss spectroscopy (EELS) measurements. Specimens are drop coated onto lacey carbon coated Cu TEM grids for TEM analysis.

**iCVD** The chemicals used for the initiated Chemical Vapor Deposition were ethylene glycol dimethacrylate (EGDMA, 98%, abcr, Germany) as the monomer and tert-butyl-peroxide (TBPO, Sigma-Aldrich, Germany) as the initiator. The reactor setup used is reported elsewhere in the literature[31]. It was evacuated by a scroll pump (nXDS 10i Edwards, Burgess Hill, UK) while the pressure was controlled by a butterfly valve (VAT 615) connected to a capacitive manometer (MKS Baratron). A Nickel Chromium (Ni80/Cr20, Goodfellow GmbH) filament array was resistively heated by a power supply (DELTA ELEKTRONIKA, SM 7020-D). The samples were placed below the filament array on a copper sample stage cooled by a circulating thermostat (Huber CC-K6). The pEGDMA coating was deposited for 40 mins by using an EGDMA monomer gas flow of 0.3 sccm and a TBPO gas flow of 0.3 sccm, a power of 42 W to heat the filament array and a pressure of 40 Pa, while the sample stage was cooled to 30°C. The thin film was also deposited on Si-wafer cut-outs resulting in a thickness of 150 nm. The film on the samples is much thinner (~5.5 nm), as the thermal conductivity of the $SiO_2$ framework structure is minimal, reducing the adsorption by the gas phase species, as well as exhibiting a large surface area in a small volume.

**Luminescence** spectra were recorded in quartz ampoules at room temperature using a Fluorolog 3 spectrometer (Horiba, Jovin Yvon GmbH, Unterhaching, Germany) equipped with an iHR-320-FA triple grating imaging spectrograph, a Syncerity charge-coupled device (CCD) detector and a 450 W Xe lamp.

**Long-term stability** was measured by placing the samples in a cuvette and adding 1 ml of water. Images were recorded under UV light (Thorlabs M375L4, 375 nm) using a Canon EOS RP with an illumination time of 1/30 s, F/22 and ISO-12800.

**Declaration of Competing Interest**

The authors declare no conflict of interest.

**Acknowledgment**

The authors acknowledge funding from the European Union's Horizon 2020 Research and Innovation Programme under grant agreement No GrapheneCore3 881603. E. E. S. Teotonio also thanks the CAPES (CAPES, grant 88887.371434/2019-00) and Federal University of Paraíba (UFPB) for financial support.

**Data availability**

The data that support the findings of this study are available from the corresponding authors upon request.

Supplementary Information

# Functional light diffusers based on hybrid CsPbBr$_3$/SiO$_2$ aero-framework structures for laser light illumination and conversion

Lena M. Saure[1], Jonas Lumma[1,2], Niklas Kohlmann[3], Torge Hartig[4], Ercules E.S. Teotonio[2,5], Shwetha Shetty[6], Narayanan Ravishankar[6], Lorenz Kienle[3,7], Franz Faupel[4], Stefan Schröder[4], Rainer Adelung[1,7], Huayna Terraschke[2,7*], Fabian Schütt[1,7*]

[1] Functional Nanomaterials, Department for Materials Science, Kiel University, Kaiser Str. 2, 24143 Kiel, Germany

[2] Institute of Inorganic Chemistry, Christian-Albrechts-Universität zu Kiel, Max-Eyth-Str. 2, D-24118 Kiel, Germany. E-Mail: hterraschke@ac.uni-kiel.de

[3] Synthesis and Real Structure, Institute for Materials Science, Kiel University, Kaiser Str. 2, 24143 Kiel, Germany

[4] Chair for Multicomponent Materials, Department for Materials Science, Kiel University, Kaiser Str. 2, 24143 Kiel, Germany

[5] Department of Chemistry, Federal University of Paraíba, 58051-970 João Pessoa, Paraíba, Brazil

[6] Materials Research Centre, Indian Institute of Science, Bangalore, India

[7] Kiel Nano, Surface and Interface Science KiNSIS, Kiel University, Christian-Albrechts-Platz 4, 24118 Kiel, Germany. E-Mail: fas@tf.uni-kiel.de

**Details on light scattering properties**

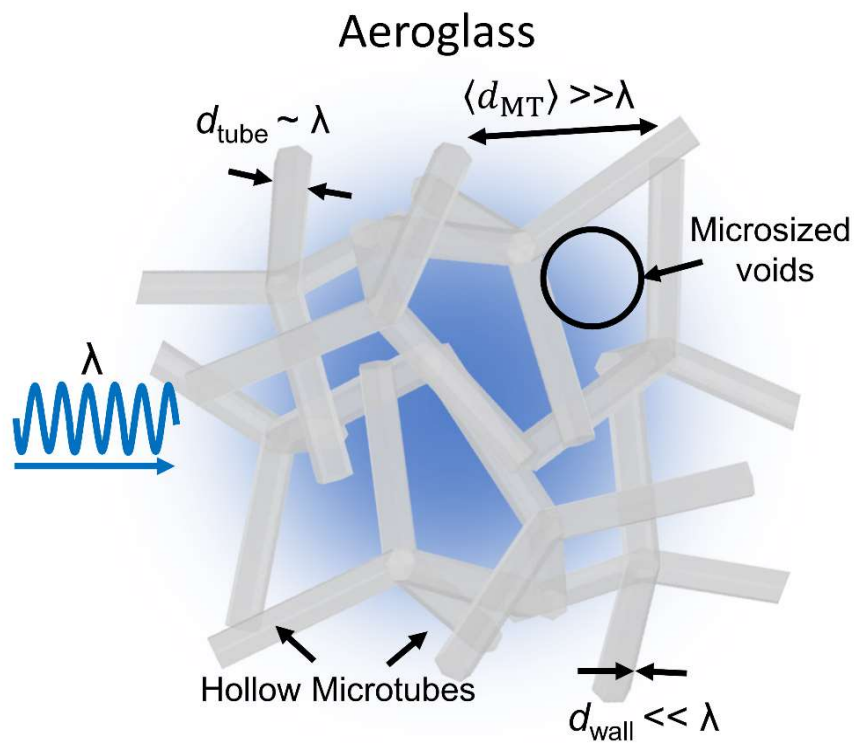

**Figure S 1** Schematic of SiO$_2$ framework structure, named aeroglass, demonstrating different feature sized compared to light in the visible spectrum.

Based on template density 0.6 g cm$^{-3}$, calculated microtube density 13.6·10$^8$ cm$^{-3}$

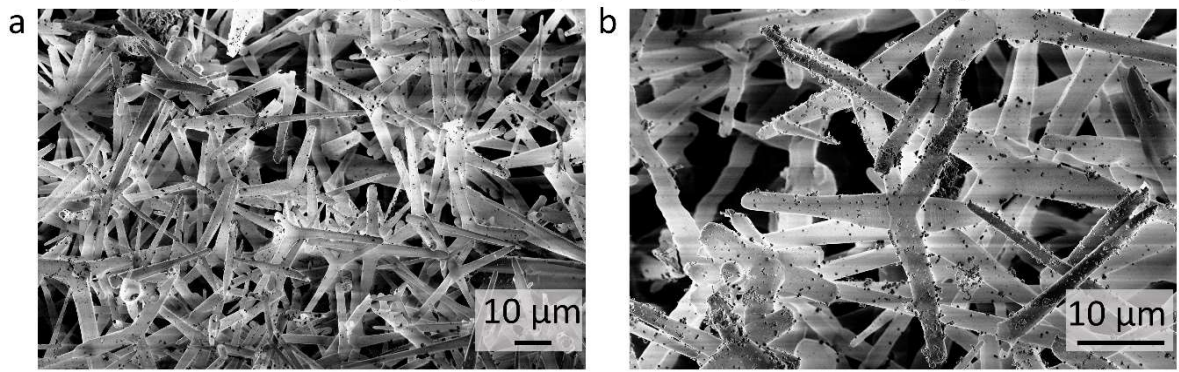

Based on template density 0.9 g cm$^{-3}$, calculated microtube density 20.4·10$^8$ cm$^{-3}$

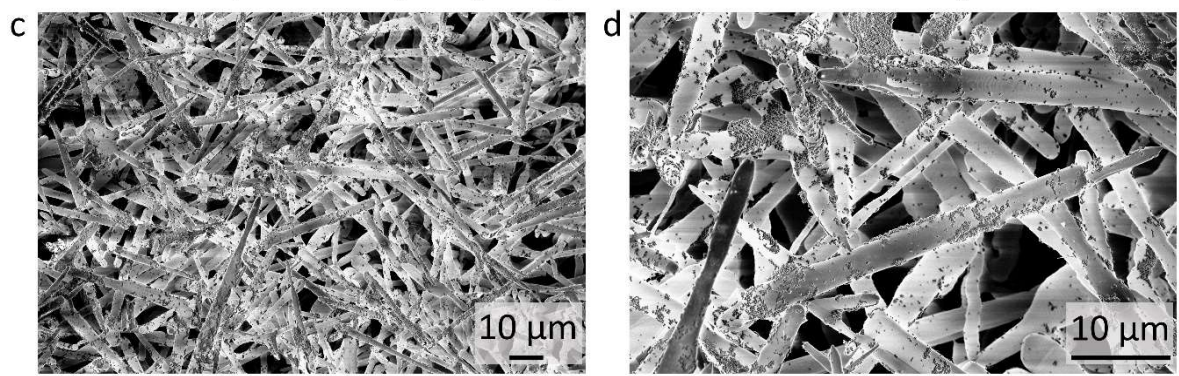

**Figure S 2** SEM images of aeroglass based on different densities of ZnO template. (a), (b) aeroglass based on template density 0.6 g cm$^{-3}$. (c), (d) aeroglass based on template density 0.9 g cm$^{-3}$.

**Supplementary Note 1 – Calculation of microtube density**

To calculate the number of hollow microtubes within a unit volume of aeroglass the number of ZnO tetrapod arms is calculated. For this, it was assumed that the ZnO tetrapods arms are rod-shaped with a diameter d = 2 µm and a length l of 25 µm and the volume of a tetrapod arm $V_{rod}$ is calculated by:

$$V_{rod} = \frac{\pi * (\frac{d}{2})^2 * l}{10^{12}} \tag{S1}$$

Knowing the template geometries, the density of the template ($\rho_{template}$ = 0.3 g cm$^{-3}$, 0.6 g cm$^{-3}$ and 0.9 g cm$^{-3}$, respectively) and the density of ZnO ($\rho_{ZnO}$ = 5.61 g cm$^{-3}$), the number of ZnO tetrapods arms $n_{rod}$ in the template can be calculated:

$$n_{rod} = \frac{V_{ZnO}}{V_{rod}} = \frac{\rho_{template}}{V_{tetrapod} * \rho_{ZnO}} \tag{S2}$$

As the number of ZnO tetrapod arms equals the number of hollow SiO$_2$ microtubes, the density of microtubes per unit volume can be calculated:

$$\rho_{microtube} = \frac{n_{rod}}{V_{template}} \tag{S3}$$

The calculation results are summarized in the **Table 1**:

**Table 1** Calculation results of microtube densities based on ZnO template density.

| ZnO template density [g cm$^{-3}$] | Density of microtubes [10$^8$ cm$^{-3}$] |
|---|---|
| 0.3 | 6.8 |
| 0.6 | 13.6 |
| 0.9 | 20.4 |

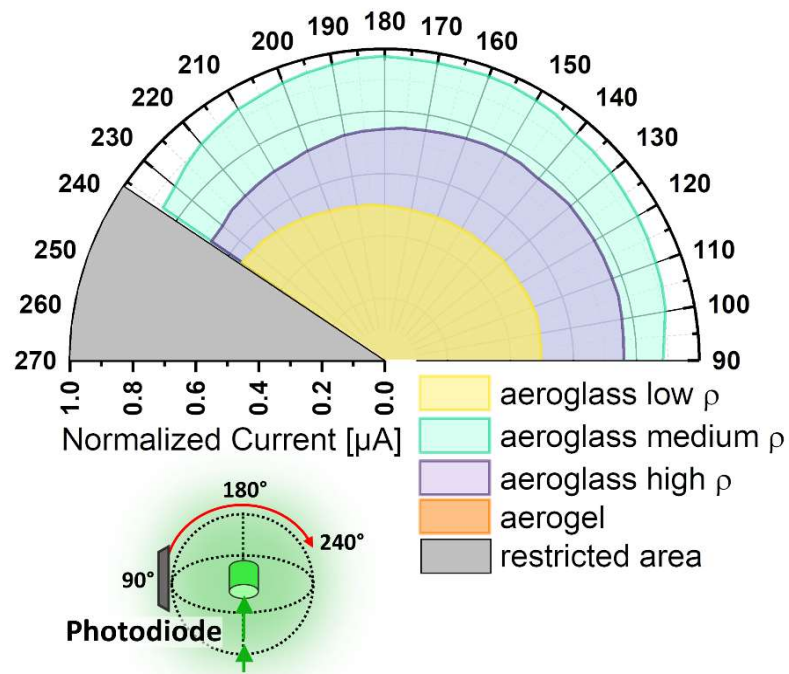

**Figure S 3** Light scattering in x-z-plane of aeroglass with different densities of scattering centers (low, medium and high), as well as of a conventional silica aerogel. The intensity of the photocurrent is normalized to the transmitted light.

**Speckle contrast determination**

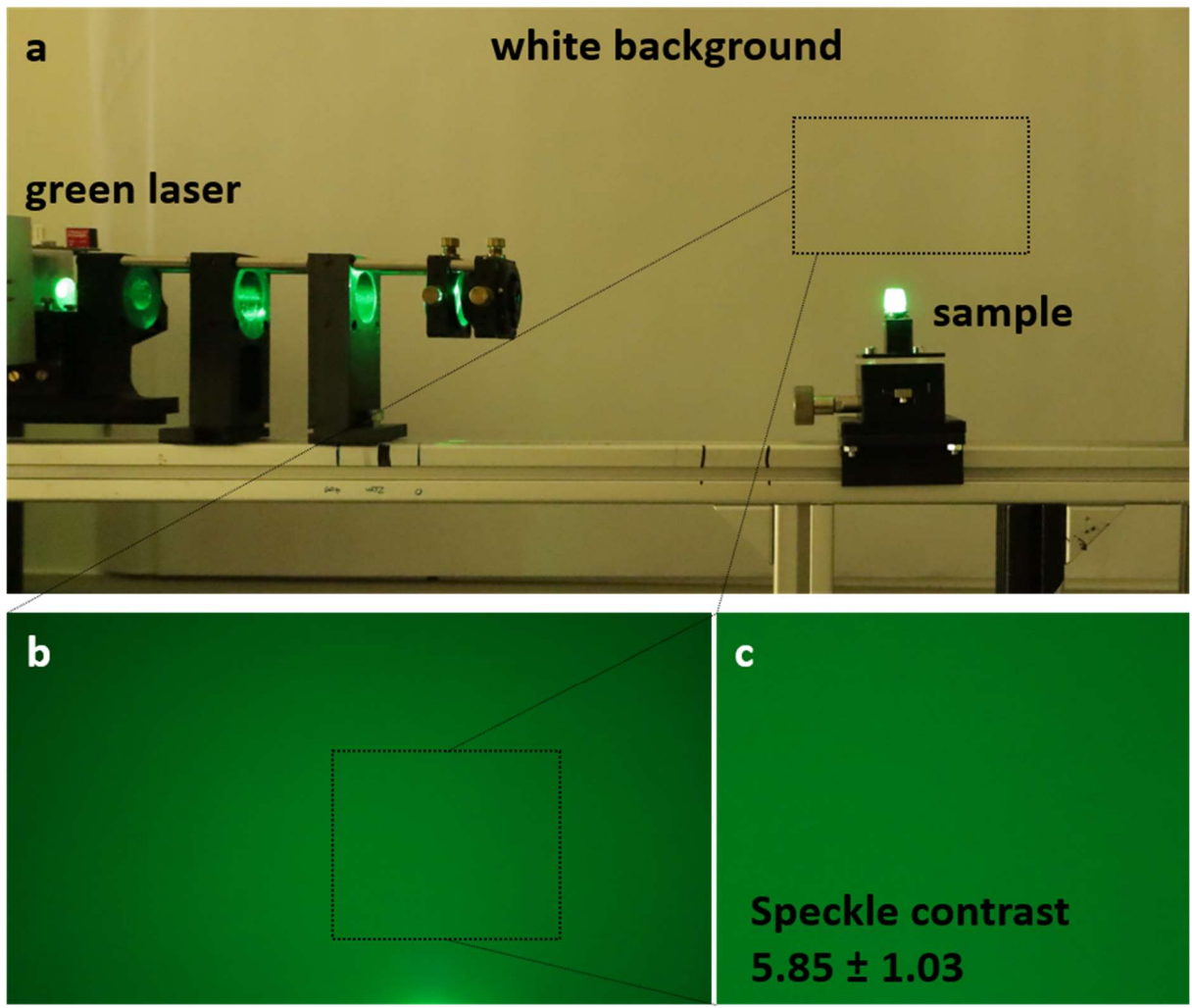

**Figure S 4** Speckle reduction of Aeroglass. (a) set-up of green laser illuminating the aeroglass sample. The square on the white background denotes the area which has been photographed and considered for speckle contrast calculations. (b) Photograph of white background illuminated by scattered light from the aeroglass sample. (c) Cutout of photograph, used for speckle contrast calculations.

**Supplementary Note 2 – Calculation of speckle contrast**

The speckle contrast of an aeroglass sample was determined using a focused laser green laser (450 nm, 100 mW) to illuminate the sample. The objective speckle pattern was evaluated from a white wall in 90° angle to the incoming light beam (distance to sample is 40 cm). The speckle pattern is captured with a photocamera (Canon EOS RP) with a focal distance of 105 mm, an exposure time of 1/60 s and an aperture of 7.1, placed slightly above the sample with a distance of ~20 cm.

From the photograph a representative area (cf. Figure S 4) is selected to determine the speckle contrast using Gatan Microscopy Suite. First, the image is converted into black and white, followed by the calculation of the mean intensity I and the standard deviation σ.

From this, the speckle contrast χ can be calculated by[1]:

$$\chi = \sigma \, I^{-1} \tag{1}$$

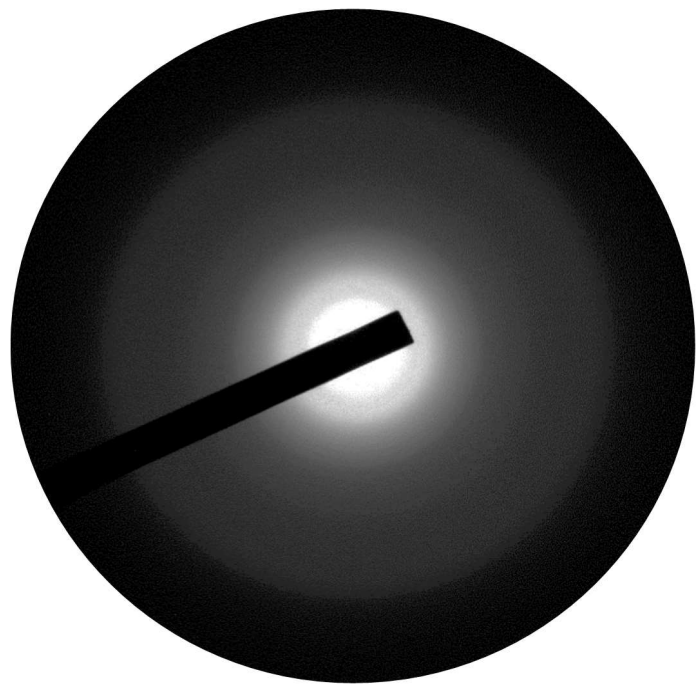

**Figure S 5** Electron diffraction of pure aeroglass, revealing amorphous structure.

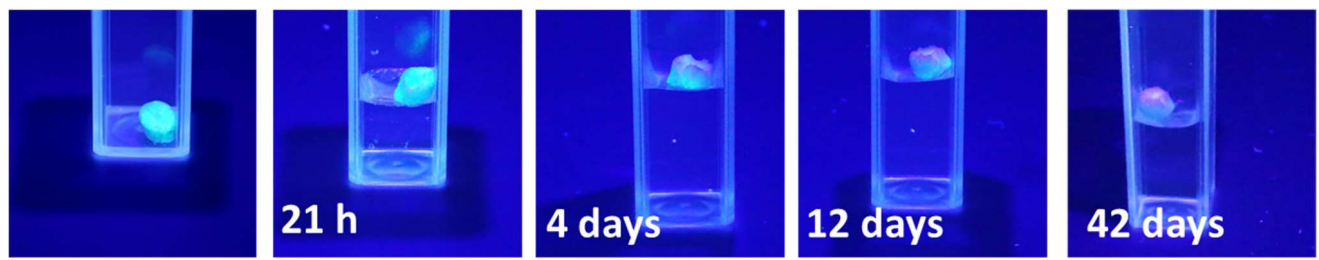

**Figure S 6** Long-term stability of aeroglass-QD-iCVD. Images were taken at different time points and show the gradual decrease in luminescence.

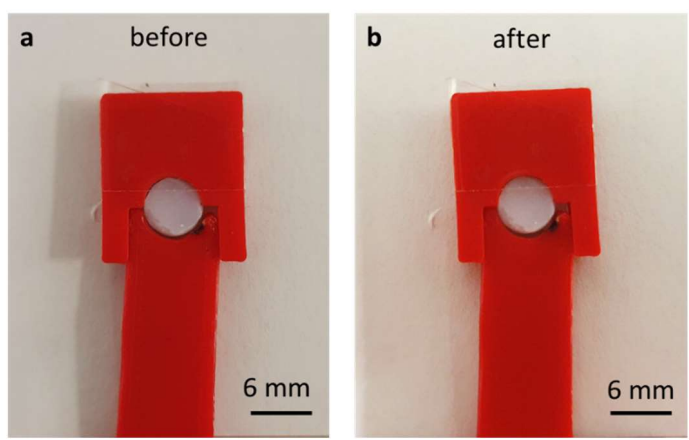

**Figure S 7** Aeroglass in sample holder (a) before and (b) after illumination with a 5W blue laser for 1 min.

**Supplementary Note 3 – Determination of layer thickness of pEGDMA coating**

The wall thickness of an individual microtube arm of a f-SiO2-QD sample as well as of a f-SiO2-QD-iCVD samples is determined by electron energy loss spectroscopy log ratio method. Performing a line scan across a hollow microtube yields the relative thickness in multiples of the inelastic electron mean free path (eMFP). The difference in eMFP between an iCVD-coated and non-coated microtube arms is 0.073 eMFP, which is the thickness of the pEGDMA coating. Calculating the mean free path of pEGDMA after Malis et al. [2] results in a mean free path of $\lambda_{EGDMA}$ ~ 150 nm, thus, the thickness of the pEGDMA coating is approximately 5.5 nm.

It has to be noted, that the relative thickness of the pEGDMA coating has to be divided by a factor of two, as the electron beam passes twice through the hollow microtube.